\documentclass[showpacs,aps,prl,twocolumn,superscriptaddress]{revtex4}
\usepackage{graphicx} %graphics handler
\usepackage{dcolumn}% Align table columns on decimal point
\usepackage{bm}% bold math
\usepackage{amssymb,amsmath}
\usepackage{epstopdf}

\begin{document}
\title{Terahertz Coherent Control of a Landau-Quantized Two-Dimensional Electron Gas}

\normalsize
\author{T. Arikawa}
\author{X. Wang}
\affiliation{Department of Electrical and Computer Engineering, Rice University, Houston, Texas 77005, USA}
%\affiliation{The Richard E.~Smalley Institute for Nanoscale Science and Technology, Rice University, Houston, Texas 77005, USA}
\affiliation{Department of Physics and Astronomy, Rice University, Houston, Texas 77005, USA}

\author{D. J. Hilton}
\affiliation{Department of Physics, University of Alabama at Birmingham, Birmingham, AL 35294, USA}

\author{J. L. Reno}
\affiliation{Center for Integrated Nanotechnologies, Sandia National Laboratories, Albuquerque, NM 87123, USA}

\author{W. Pan}
\affiliation{Sandia National Laboratories, Albuquerque, NM 87123, USA}

\author{J.~Kono}
\email[]{kono@rice.edu}
\thanks{corresponding author.}
\affiliation{Department of Electrical and Computer Engineering, Rice University, Houston, Texas 77005, USA}
%\affiliation{The Richard E.~Smalley Institute for Nanoscale Science and Technology, Rice University, Houston, Texas 77005, USA}
\affiliation{Department of Physics and Astronomy, Rice University, Houston, Texas 77005, USA}

\date{\today}

\begin{abstract}
We demonstrate coherent control of cyclotron resonance (CR) in a two-dimensional electron gas (2DEG).  We use a sequence of terahertz pulses to control the amplitude of CR oscillations in an arbitrary fashion via phase-dependent coherent interactions.  We observe a self-interaction effect, where the 2DEG interacts with the terahertz field emitted by itself within the decoherence time, resulting in a revival and collapse of quantum coherence.  These observations are accurately describable using {\em single-particle} optical Bloch equations, showing no signatures of electron-electron interactions, which verifies the validity of Kohn's theorem for CR in the coherent regime.
\end{abstract}

%\keywords{}

\maketitle


%Quantum coherence is at the heart of modern condensed matter physics as well as emerging technologies such as quantum computation.	In single-electron systems, e.g., atoms and quantum dots, coherent light-matter interaction1 has been well studied and provides powerful means for manipulating qubits2.	However, much room remains for further studies in many-electron systems, where electron-electron interactions play a decisive role in their quantum-coherent dynamics3-5. Here, 

Quantum coherence and many-body effects are at the heart of modern condensed matter physics as well as solid-state quantum technologies.  Since the advent of femtosecond lasers, coherent dynamics in solids have attracted much attention due to their strikingly different behaviors from those of atoms~\cite{ChemlaShah01Nature,AxtKuhn04RPP}.  These drastic differences between solids and atoms are direct consequences of many-body interactions unique to condensed matter, enhanced by other solid-state ingredients such as carrier-phonon interactions, localization, and quantum confinement. In addition, novel device possibilities such as topological quantum computation utilizing unconventional many-body properties~\cite{SarmaetAl05PRL} have stimulated significant interest in coherent carrier dynamics in solids.  Much success exists in manipulating single-particle systems such as quantum dots, whereas manipulation of many-particle quantum states has been an elusive goal.

Here, we successfully demonstrate coherent control of a high-density two-dimensional electron gas (2DEG) in a quantizing magnetic field via cyclotron resonance (CR), i.e., inter-Landau-level transitions.  We used terahertz (THz) time-domain spectroscopy~\cite{Mittleman02Book,Tonouchi07NaturePhoton}, which can measure the amplitude and phase of THz fields~\cite{KerstingetAl00OL,LuoetAl04PRL,KrolletAl07Nature,GunteretAl09Nature} as well as controlling coherent dynamics using multiple pulses~\cite{ColeetAl01Nature,YanoetAl09JJAP,KampfrathetAl11NaturePhoton}.  Strikingly, our results indicate that this many-body system behaves as a ``two-level atom''~\cite{AllenEberly87Book}, strictly obeying single-particle optical Bloch equations.  We attribute this behavior to Kohn's theorem regarding the absence of many-body effects in CR~\cite{Kohn61PR}, which has been proven through many continuous-wave experiments using incoherent spectroscopies~\cite{McCombeWagner75Review1,Kono01MMR}.  Our results confirm its applicability to the coherent regime, opening up new opportunities for probing and controlling coherent dynamics in quantum Hall systems.

The sample we studied contained a modulation-doped GaAs single quantum well with an electron density of 2.0 $\times$ 10$^{11}$~cm$^{-2}$ and a 4-K mobility of 3.7 $\times$ 10$^6$~cm$^2$/Vs.	  Our THz magneto-spectroscopy system~\cite{WangetAl07OL,Suppl} is schematically shown in Fig.~1(a).  Coherent THz pulses were generated from a nitrogen gas plasma~\cite{ThomsonetAl07LPR,KarpowiczetAl09JMO} created by focusing both the fundamental and second harmonic of the output of a chirped-pulse amplifier (CPA-2001, Clark-MXR, Inc.).  The first wire-grid polarizer (WGP) was placed before focusing the THz wave to the sample to make the incident wave linearly polarized along the $x$-axis.  The transmitted THz electric field through the sample was directly measured via electro-optic sampling using a (110) ZnTe crystal.  To measure the $y$-component of the transmitted THz wave, the second WGP was placed in cross-Nicole geometry before the ZnTe crystal, whose [001] axis was perpendicular to the probe polarization ($y$-polarized).  The $x$-component was measured by setting the second WGP parallel to the first WGP and the [001] direction of the ZnTe crystal parallel to the probe polarization.  The two geometries of electro-optic sampling had the same sensitivity to the $y$- and $x$-components of the THz electric field, respectively, enabling quantitative comparison with each other~\cite{Suppl,PlankenetAl01JOSAB}.  The frequency bandwidth was from 0.3~THz to 2.6~THz.  The sample was cooled down to 1.4~K in an optical cryostat with a superconducting magnet.

%%%%%  FIG. 1  %%%%%
\begin{figure}
\includegraphics[scale=0.93]{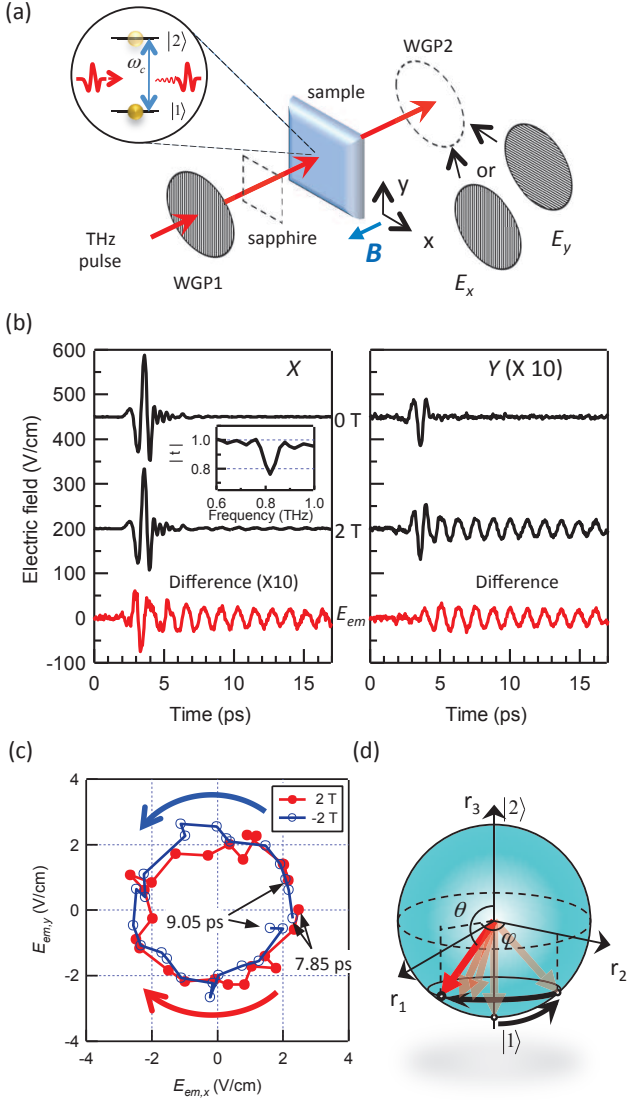}
\caption{(color online) (a)~Schematic of the experimental setup. WGP: wire grid polarizer. The incident THz pulses are linearly ($x$-) polarized. (b)~Transmitted THz pulse at 0 and 2~T and the re-emitted THz wave, $E_{\rm em}$, measured with WGP2 parallel (X, left) and perpendicular (Y, right) to WGP1. Traces are vertically offset for clarity. Inset: amplitude ratio $|t|$ in the frequency domain showing CR as a dip.  (c)~Parametric plot showing the CR active mode absorption at $\pm$2~T.  (d)~Bloch sphere for coherent CR.}
\label{CCR}
\end{figure}
%%%%%%%%%%%%%%%%%%%%%

Figure 1(b) shows coherent CR oscillations in the time domain. The left graph shows the $x$-component of the transmitted THz field with [$E(B)$] and without [$E(0)$] a magnetic field ($B$) of 2~T.  Small-amplitude oscillations appear after the original THz pulse when $B$ is present. To extract the $B$-induced oscillations, the 0-T waveform was subtracted from the 2-T waveform, producing a decaying sinusoid with a cyclotron frequency $\omega_c/2\pi = eB/2\pi m^*$ of 0.816~THz, where $e$ is the electronic charge, $m^* = 0.068m_0$~\cite{WangetAl07OL}, and $m_0 = 9.11 \times 10^{-31}$~kg.  The inset shows the amplitude ratio between with and without $B$ in the frequency domain, showing a dip at $\omega_c/2\pi$.  The right graph in Fig.~1(b) shows the $y$-component of the transmitted THz field ($\times$ 10).  Again, clear CR oscillations are seen when $B$ is present, also a sinusoid with the same CR frequency, but with a different phase from the $x$-component. The appearance of $B$-induced oscillations in the $y$-component comes from the finite off-diagonal elements of the magneto-optical conductivity tensor~\cite{PalikFurdyna70RPP}.  Figure 1(c) shows a parametric plot of these oscillations from 7.85~ps to 9.05~ps, which approximately corresponds to one full period of cyclotron oscillation (1/0.816~THz = 1.22~ps).  Each point represents the tip of the electric field vector at each time, which rotates clockwise as time progresses and forms a closed circle.  This clearly shows that the absorbed THz wave is the circularly-polarized CR active mode;  the direction of rotation is reversed when the direction of $B$ is reversed (-2~T).

Let us now turn to a quantum mechanical treatment of the observed coherent oscillations. The energy of a 2DEG in a perpendicular $B$ is quantized into Landau levels (LLs) separated by $\hbar\omega_c$.  At 2~T, the LL filling factor is 4.1, i.e., the lowest-two LLs ($|0\rangle$ and $|1\rangle$) are completely filled (including the spin degeneracy), the next level $|2\rangle$ is almost empty, and all higher levels are empty.  Inter-LL transitions are allowed only between adjacent levels~\cite{McCombeWagner75Review1,Kono01MMR}, and hence, we treat the system as a two-level system ($|1\rangle$ and $|2\rangle$) as shown in Fig.~1(a).  The incident THz pulse creates a coherent superposition, $|\psi\rangle = C_1 \exp(-iE_1t/\hbar)|1\rangle + C_2 \exp(-iE_2t/\hbar)|2\rangle$, where $C_{i = 1,2}$ are probability amplitudes satisfying $|C_1|^2 + |C_2|^2= 1$.  Alternatively, $|\psi\rangle$ can be expressed as a Bloch vector using coordinates ($\theta$,$\varphi$) as $|\psi\rangle = \sin(\theta/2)|1\rangle + e^{i\varphi}\cos(\theta/2)|2\rangle$ [see Fig.~1(d)]. When the system is initially in the ground state ($|1\rangle$), the tip of the vector is at ($\theta$,$\varphi$) = ($\pi$, undefined).  The tip will move to ($\theta$,$\varphi$) = ($\theta_1$,0) by a unitary rotation about the $r_1$ axis through coherent interaction with the incident THz pulse.  The change in the polar angle $\pi - \theta_1$ is equal to the THz pulse area and is small (around 0.1 degrees) in our experiments due to the weak fields employed.  After that, the Bloch vector starts rotating at frequency $\omega_c$, acquiring a dynamical phase $\varphi(t)$.  It has an oscillating dipole moment $\langle \mu \rangle = 2\mu_{12} {\rm Re}[C_1^* C_2 \exp(-i\omega_c t)]$, which re-emits a THz wave $E_{em}$ at $\omega_c$.  Here, $\mu_{12} = \langle 1|\hat{\mu}|2\rangle$ is the dipole matrix element between the states $|1\rangle$ and $|2\rangle$.  In this picture, CR is the destructive interference of the incident field $E(0)$ and the re-emitted field $E_{em}$, i.e., $E(B) = E(0) + E_{em}$. Thus, the $B$-induced oscillations [bottom traces in Fig.~1(b)] are the re-emitted THz wave, $E_{em} = E(B) - E(0)$, due to the induced THz dipole.

%%%%%  FIG. 2  %%%%%
\begin{figure}
\includegraphics[scale=0.92]{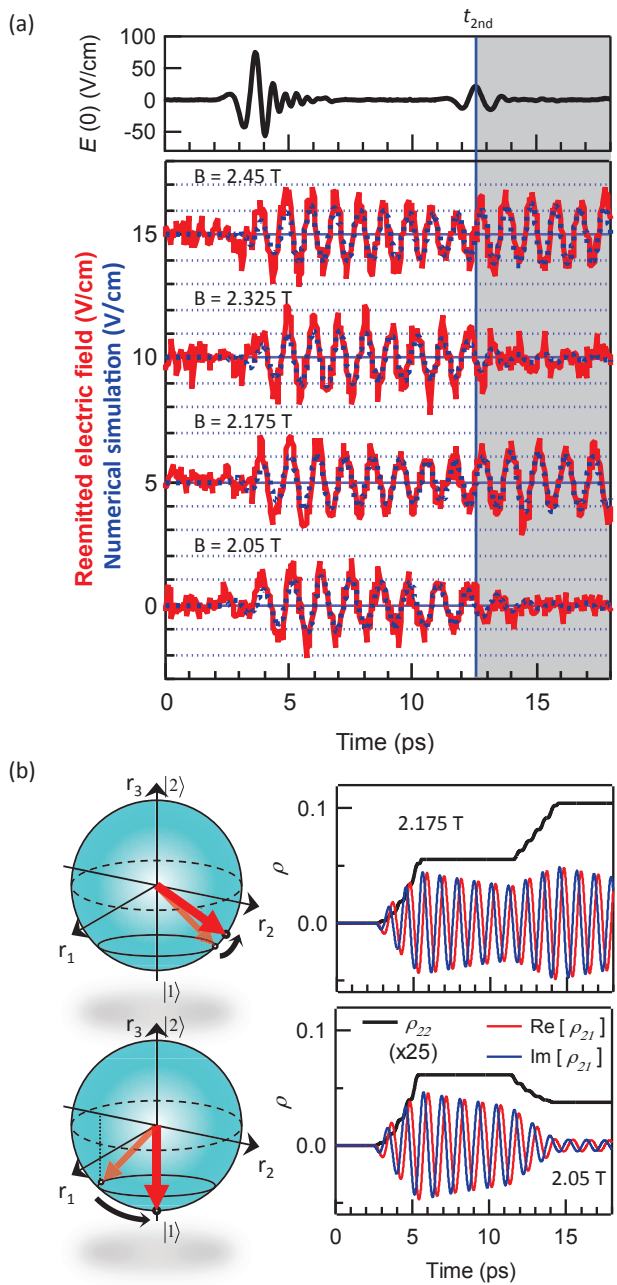}
\caption{(color online) Coherent control of CR. (a)~Incident (upper) and re-emitted field (lower) at several magnetic fields. The second pulse was created through the internal reflection inside a 430~$\mu$m-thick sapphire plate [Fig.~1(a)].  (b)~Bloch vector dynamics and calculated density matrix elements.}
\label{control}
\end{figure}
%%%%%%%%%%%%%%%%%%%%%

Next, we demonstrate coherent control of CR using a sequence of two THz pulses. Figure 2(a) shows incident pulses $E(0)$ (upper) and the $y$-component of the re-emitted field (lower) at several magnetic fields.  After the second pulse is incident, re-emission is quenched at 2.05~T and 2.325~T, whereas re-emission is enhanced at 2.175~T and 2.45~T.  The difference between these two cases is the oscillation {\em phase} at the arrival time $t_{\rm 2nd}$ of the second pulse.  At 2.05~T, the sinusoid starting from 3.6~ps exhibits 7.5 periods by $t_{\rm 2nd}$, i.e., the phase is an odd-integer multiple of $\pi$; in this case, re-emission is quenched.  In contrast, at 2.175~T where re-emission is enhanced, there are 8 periods, i.e., the phase is an even-integer multiple of $\pi$. The same is true for 2.325~T and 2.45~T.  These situations are again understandable using the Bloch sphere [Fig.~2(b)].  In the quenching case, the phase of an odd-integer multiple of $\pi$ at $t_{\rm 2nd}$ means that the Bloch vector is at ($\theta_1$,$\pi$) and the rotation operation about the $r_1$ axis by the second THz pulse moves the vector back to $|1\rangle$ [Fig.~2(b), lower].  In the enhancement case, at $t_{\rm 2nd}$ the Bloch vector is at ($\theta_1$,0), and the rotation operation further increases $\theta$.

Coherent dynamics of two-level atoms are describable through the optical Bloch equations~\cite{AllenEberly87Book,Boyd08Book}.  We applied this theoretical framework to the response of the many-electron 2DEG simply by treating the system as a collection of independent two-level systems (i.e., neglecting electron-electron interactions)~\cite{Suppl}.  The right graphs in Fig.~2(b) show the time evolution of the density matrix for 2.05~T (lower) and 2.175~T (upper), corresponding to the quenching and enhancing cases, respectively.  At 2.05~T, the upper state population, $\rho_{22}$, increases through interaction with the first pulse (absorption), but decreases through interaction with the second pulse (stimulated emission). As expected, the population $\rho_{22}$ is small due to the weak THz pulses we employed, and only the beginning of Rabi oscillations is seen.  The coherence $\rho_{21}$ is also decreased by the second pulse.  In contrast, at 2.175~T, both the population and coherence are increased by the second pulse. The calculated re-emitted waves are plotted in Fig.~2(a) (dashed blue curves), reproducing the experimental data well.  There is no adjustable fitting parameters in these simulations~\cite{Suppl}.

%%%%%  FIG. 3  %%%%%
\begin{figure}
\includegraphics [scale=0.95] {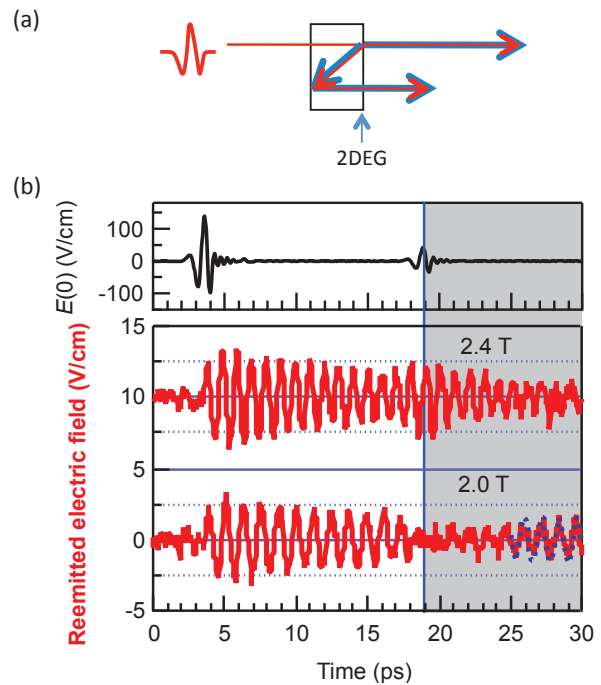}
\caption{(color online) (a)~The interaction of the 2DEG with its self-re-emitted wave. Thick blue arrows: re-emitted THz wave from the 2DEG; thin red arrows: reflected and transmitted incident THz pulse. (b)~The incident electric field (upper) and re-emitted field (lower). The second incident THz pulse is due to the internal reflection inside the sample substrate. The blue dashed curve is an extended fitting curve to the initial oscillation from 5~ps to 17~ps with a sine wave.}
\label{self}
\end{figure}
%%%%%%%%%%%%%%%%%%%%%

Finally, we describe an unusual situation where the 2DEG interacts with the THz field emitted by itself within the decoherence time [Fig.~3(a)].  This coherent `self-interaction' resulted in a revival and collapse of quantum coherence.  Figure 3(b) shows a re-emitted THz wave recorded for a longer time (without any intentionally created second pulse). The THz pulse reflected back to the substrate by the 2DEG acquires trailing oscillations due to the backward CR re-emission by the 2DEG (similar to the transmitted THz wave). This THz pulse is further reflected by the back surface of the substrate and incident {\em again} onto the 2DEG at 19~ps (still less than the decoherence time).  As a result, at 2~T, the amplitude of re-emission, once quenched by the incident reflection THz pulse, grows again with time because of the interaction with the long-lasting self-re-emitted coherent THz wave.  Note that the revived oscillation is in-phase with the original one in this case (blue dashed curve).  In contrast, at 2.4~T, the re-emission oscillations, initially enhanced by the reflection THz pulse, decays quickly due to interaction with the self-re-emitted coherent THz wave.  The difference between the two cases comes from the same mechanism discussed above, i.e., $\pi$ difference in the arrival phase, which is again described well via optical Bloch equations.

The agreement between experiment and theory highlights the striking fact that a Landau-quantized 2DEG in a weak coherent THz field behaves in exactly the same manner as a two-level atom in a coherent radiation field.  We interpret this behavior as a result of Kohn's theorem~\cite{Kohn61PR} regarding the absence of an influence of electron-electron interactions on CR frequencies.  However, it is important to note that our results obtained through a coherent time-domain method reveals more about coherent CR dynamics than what Kohn's theorem predicts. Namely, our results show that not only the CR frequency but also the non-equilibrium dynamical response of a 2DEG under multiple-pulse irradiation is not affected by electron-electron interactions.  This novel result thereby broadens the realm of applicability of Kohn's theorem.

It should be also noted that the CR dephasing mechanism is not understood.  At low temperatures where phonons are suppressed, the main cause of dephasing is not clear.  Defects might ultimately limit the decoherence time as in DC transport (momentum relaxation time $\sim$ 143~ps in our sample).  The dephasing of {\em interband} photo-excited carriers (for which Kohn's theorem does not apply) is strongly dependent on the filling factor~\cite{KaradimitriouetAl10PRB,BroocksetAl08PSSB}.  In contrast, in the present {\em intraband} excitation experiments, no obvious filling-factor dependence was observed, which implies that dephasing due to electron-electron interactions is also prohibited by Kohn's theorem.

In summary, using time-domain THz techniques, we performed quantum control experiments on CR in a high-mobility 2DEG.  We used single-particle optical Bloch equations to successfully simulate our experimental data, highlighting the striking fact that an interacting many-electron state behaves as a single particle (Kohn's theorem for CR).  However, the current experiments were performed in the weak excitation regime -- the THz field strengths used were far from what is needed to cause Rabi oscillations.  A strong THz wave would excite electrons through equally-spaced higher-energy LLs in a cascading fashion, invalidating the two-level approximation. Nonparabolic systems such as InAs or graphene with non-equidistance LLs together with a narrowband THz wave can be used to circumvent this problem.

\begin{acknowledgments}
We thank Alexey Belyanin and Adilet Imambekov for valuable discussions.  This work was supported by the Sandia National Laboratories (Award No.~896506), the National Science Foundation (Grant No.~OISE-0530220), the Texas Higher Education Coordinating Board (Award No.~01905), and the Division of Materials Sciences and Engineering, Office of Basic Energy Sciences, U.S.~Department of Energy (Award No.~896506 at Sandia National Laboratories). This work was also performed in part at the U.S.~Department of Energy, Center for Integrated Nanotechnologies at Sandia National Laboratories (Contract DE-AC04-94AL85000).
\end{acknowledgments}


%\bibliography{TakashiCCR,jun}

\begin{thebibliography}{25}
\expandafter\ifx\csname natexlab\endcsname\relax\def\natexlab#1{#1}\fi
\expandafter\ifx\csname bibnamefont\endcsname\relax
  \def\bibnamefont#1{#1}\fi
\expandafter\ifx\csname bibfnamefont\endcsname\relax
  \def\bibfnamefont#1{#1}\fi
\expandafter\ifx\csname citenamefont\endcsname\relax
  \def\citenamefont#1{#1}\fi
\expandafter\ifx\csname url\endcsname\relax
  \def\url#1{\texttt{#1}}\fi
\expandafter\ifx\csname urlprefix\endcsname\relax\def\urlprefix{URL }\fi
\providecommand{\bibinfo}[2]{#2}
\providecommand{\eprint}[2][]{\url{#2}}

\bibitem[{\citenamefont{Chemla and Shah}(2001)}]{ChemlaShah01Nature}
\bibinfo{author}{\bibfnamefont{D.~S.} \bibnamefont{Chemla}} \bibnamefont{and}
  \bibinfo{author}{\bibfnamefont{J.}~\bibnamefont{Shah}},
  \bibinfo{journal}{Nature} \textbf{\bibinfo{volume}{411}},
  \bibinfo{pages}{549} (\bibinfo{year}{2001}).

\bibitem[{\citenamefont{Axt and Kuhn}(2004)}]{AxtKuhn04RPP}
\bibinfo{author}{\bibfnamefont{V.~M.} \bibnamefont{Axt}} \bibnamefont{and}
  \bibinfo{author}{\bibfnamefont{T.}~\bibnamefont{Kuhn}},
  \bibinfo{journal}{Rep. Prog. Phys.} \textbf{\bibinfo{volume}{67}},
  \bibinfo{pages}{433} (\bibinfo{year}{2004}).

\bibitem[{\citenamefont{Das~Sarma et~al.}(2005)\citenamefont{Das~Sarma,
  Freedman, and Nayak}}]{SarmaetAl05PRL}
\bibinfo{author}{\bibfnamefont{S.}~\bibnamefont{Das~Sarma}},
  \bibinfo{author}{\bibfnamefont{M.}~\bibnamefont{Freedman}}, \bibnamefont{and}
  \bibinfo{author}{\bibfnamefont{C.}~\bibnamefont{Nayak}},
  \bibinfo{journal}{Phys. Rev. Lett.} \textbf{\bibinfo{volume}{94}},
  \bibinfo{pages}{166802} (\bibinfo{year}{2005}).

\bibitem[{\citenamefont{Mittleman}(2002)}]{Mittleman02Book}
\bibinfo{editor}{\bibfnamefont{D.}~\bibnamefont{Mittleman}}, ed.,
  \emph{\bibinfo{title}{Sensing with Terahertz Radiation}}
  (\bibinfo{publisher}{Springer}, \bibinfo{address}{Berlin},
  \bibinfo{year}{2002}).

\bibitem[{\citenamefont{Tonouchi}(2007)}]{Tonouchi07NaturePhoton}
\bibinfo{author}{\bibfnamefont{M.}~\bibnamefont{Tonouchi}},
  \bibinfo{journal}{Nature Photon.} \textbf{\bibinfo{volume}{1}},
  \bibinfo{pages}{97} (\bibinfo{year}{2007}).

\bibitem[{\citenamefont{Kersting et~al.}(2000)\citenamefont{Kersting,
  Bratschitsch, Strasser, Unterrainer, and Heyman}}]{KerstingetAl00OL}
\bibinfo{author}{\bibfnamefont{R.}~\bibnamefont{Kersting}} {\it et~al}.,
  %\bibinfo{author}{\bibfnamefont{R.}~\bibnamefont{Bratschitsch}},
  %\bibinfo{author}{\bibfnamefont{G.}~\bibnamefont{Strasser}},
  %\bibinfo{author}{\bibfnamefont{K.}~\bibnamefont{Unterrainer}},
  %\bibnamefont{and} \bibinfo{author}{\bibfnamefont{J.~N.}
  %\bibnamefont{Heyman}}, 
  \bibinfo{journal}{Opt. Lett.}
  \textbf{\bibinfo{volume}{25}}, \bibinfo{pages}{272} (\bibinfo{year}{2000}).

\bibitem[{\citenamefont{Luo et~al.}(2004)\citenamefont{Luo, Reimann, Woerner,
  Elsaesser, Hey, and Ploog}}]{LuoetAl04PRL}
\bibinfo{author}{\bibfnamefont{C.~W.} \bibnamefont{Luo}} {\it et~al}.,
  %\bibinfo{author}{\bibfnamefont{K.}~\bibnamefont{Reimann}},
  %\bibinfo{author}{\bibfnamefont{M.}~\bibnamefont{Woerner}},
  %\bibinfo{author}{\bibfnamefont{T.}~\bibnamefont{Elsaesser}},
  %\bibinfo{author}{\bibfnamefont{R.}~\bibnamefont{Hey}}, \bibnamefont{and}
  %\bibinfo{author}{\bibfnamefont{K.~H.} \bibnamefont{Ploog}},
  \bibinfo{journal}{Phys. Rev. Lett.} \textbf{\bibinfo{volume}{92}},
  \bibinfo{pages}{047402} (\bibinfo{year}{2004}).

\bibitem[{\citenamefont{Kr\"{o}ll et~al.}(2007)\citenamefont{Kr\"{o}ll, Darmo,
  Dhillon, Marcadet, Calligaro, Sirtori, and Unterrainer}}]{KrolletAl07Nature}
\bibinfo{author}{\bibfnamefont{J.}~\bibnamefont{Kr\"{o}ll}} {\it et~al}.,
  %\bibinfo{author}{\bibfnamefont{J.}~\bibnamefont{Darmo}},
  %\bibinfo{author}{\bibfnamefont{S.~S.} \bibnamefont{Dhillon}},
  %\bibinfo{author}{\bibfnamefont{X.}~\bibnamefont{Marcadet}},
  %\bibinfo{author}{\bibfnamefont{M.}~\bibnamefont{Calligaro}},
  %\bibinfo{author}{\bibfnamefont{C.}~\bibnamefont{Sirtori}}, \bibnamefont{and}
  %\bibinfo{author}{\bibfnamefont{K.}~\bibnamefont{Unterrainer}},
  \bibinfo{journal}{Nature} \textbf{\bibinfo{volume}{449}},
  \bibinfo{pages}{698} (\bibinfo{year}{2007}).

\bibitem[{\citenamefont{G\"{u}nter et~al.}(2009)\citenamefont{G\"{u}nter,
  Anappara, Hees, Sell, Biasiol, Sorba, De~Liberato, Ciuti, Tredicucci,
  Leitenstorfer et~al.}}]{GunteretAl09Nature}
\bibinfo{author}{\bibfnamefont{G.}~\bibnamefont{G\"{u}nter}} {\it et~al}.,
  %\bibinfo{author}{\bibfnamefont{A.~A.} \bibnamefont{Anappara}},
  %\bibinfo{author}{\bibfnamefont{J.}~\bibnamefont{Hees}},
  %\bibinfo{author}{\bibfnamefont{A.}~\bibnamefont{Sell}},
  %\bibinfo{author}{\bibfnamefont{G.}~\bibnamefont{Biasiol}},
  %\bibinfo{author}{\bibfnamefont{L.}~\bibnamefont{Sorba}},
  %\bibinfo{author}{\bibfnamefont{S.}~\bibnamefont{De~Liberato}},
  %\bibinfo{author}{\bibfnamefont{C.}~\bibnamefont{Ciuti}},
  %\bibinfo{author}{\bibfnamefont{A.}~\bibnamefont{Tredicucci}},
  %\bibinfo{author}{\bibfnamefont{A.}~\bibnamefont{Leitenstorfer}},
  %\bibnamefont{et~al.}, 
  \bibinfo{journal}{Nature}
  \textbf{\bibinfo{volume}{458}}, \bibinfo{pages}{178} (\bibinfo{year}{2009}).

\bibitem[{\citenamefont{Cole et~al.}(2001)\citenamefont{Cole, Williams, King,
  Sherwin, and Stanley}}]{ColeetAl01Nature}
\bibinfo{author}{\bibfnamefont{B.~E.} \bibnamefont{Cole}} {\it et~al}.,
  %\bibinfo{author}{\bibfnamefont{J.~B.} \bibnamefont{Williams}},
  %\bibinfo{author}{\bibfnamefont{B.~T.} \bibnamefont{King}},
  %\bibinfo{author}{\bibfnamefont{M.~S.} \bibnamefont{Sherwin}},
  %\bibnamefont{and} \bibinfo{author}{\bibfnamefont{C.~R.}
  %\bibnamefont{Stanley}}, 
  \bibinfo{journal}{Nature}
  \textbf{\bibinfo{volume}{410}}, \bibinfo{pages}{60} (\bibinfo{year}{2001}).

\bibitem[{\citenamefont{Yano et~al.}(2009)\citenamefont{Yano, Nakagawa, and
  Shinojima}}]{YanoetAl09JJAP}
\bibinfo{author}{\bibfnamefont{R.}~\bibnamefont{Yano}} {\it et~al}.,
  %\bibinfo{author}{\bibfnamefont{K.}~\bibnamefont{Nakagawa}}, \bibnamefont{and}
  %\bibinfo{author}{\bibfnamefont{H.}~\bibnamefont{Shinojima}},
  \bibinfo{journal}{Jpn. J. Appl. Phys.} \textbf{\bibinfo{volume}{48}},
  \bibinfo{pages}{022401} (\bibinfo{year}{2009}).

\bibitem[{\citenamefont{Kampfrath et~al.}(2011)\citenamefont{Kampfrath, Sell,
  Klatt, Pashkin, M{\"a}hrlein, Dekorsy, Wolf, Fiebig, Leitenstorfer, and
  Huber}}]{KampfrathetAl11NaturePhoton}
\bibinfo{author}{\bibfnamefont{T.}~\bibnamefont{Kampfrath}} {\it et~al}.,
  %\bibinfo{author}{\bibfnamefont{A.}~\bibnamefont{Sell}},
  %\bibinfo{author}{\bibfnamefont{G.}~\bibnamefont{Klatt}},
  %\bibinfo{author}{\bibfnamefont{A.}~\bibnamefont{Pashkin}},
  %\bibinfo{author}{\bibfnamefont{S.}~\bibnamefont{M{\"a}hrlein}},
  %\bibinfo{author}{\bibfnamefont{T.}~\bibnamefont{Dekorsy}},
  %\bibinfo{author}{\bibfnamefont{M.}~\bibnamefont{Wolf}},
  %\bibinfo{author}{\bibfnamefont{M.}~\bibnamefont{Fiebig}},
  %\bibinfo{author}{\bibfnamefont{A.}~\bibnamefont{Leitenstorfer}},
  %\bibnamefont{and} \bibinfo{author}{\bibfnamefont{R.}~\bibnamefont{Huber}},
  \bibinfo{journal}{Nature Photon.} \textbf{\bibinfo{volume}{5}},
  \bibinfo{pages}{31} (\bibinfo{year}{2011}).

\bibitem[{\citenamefont{Allen and Eberly}(1987)}]{AllenEberly87Book}
\bibinfo{author}{\bibfnamefont{L.}~\bibnamefont{Allen}} \bibnamefont{and}
  \bibinfo{author}{\bibfnamefont{J.~H.} \bibnamefont{Eberly}},
  \emph{\bibinfo{title}{Optical Resonance and Two-level Atoms}}
  (\bibinfo{publisher}{Dover}, \bibinfo{address}{New York},
  \bibinfo{year}{1987}).

\bibitem[{\citenamefont{Kohn}(1961)}]{Kohn61PR}
\bibinfo{author}{\bibfnamefont{W.}~\bibnamefont{Kohn}}, \bibinfo{journal}{Phys.
  Rev.} \textbf{\bibinfo{volume}{123}}, \bibinfo{pages}{1242}
  (\bibinfo{year}{1961}).

\bibitem[{\citenamefont{McCombe and Wagner}(1975)}]{McCombeWagner75Review1}
\bibinfo{author}{\bibfnamefont{B.~D.} \bibnamefont{McCombe}} \bibnamefont{and}
  \bibinfo{author}{\bibfnamefont{R.~J.} \bibnamefont{Wagner}}, in
  \emph{\bibinfo{booktitle}{Advances in Electronics and Electron Physics}},
  edited by \bibinfo{editor}{\bibfnamefont{L.}~\bibnamefont{Marton}}
  (\bibinfo{publisher}{Academic Press}, \bibinfo{address}{New York},
  \bibinfo{year}{1975}), vol.~\bibinfo{volume}{37}, pp. \bibinfo{pages}{1--78}.

\bibitem[{\citenamefont{Kono}(2001)}]{Kono01MMR}
\bibinfo{author}{\bibfnamefont{J.}~\bibnamefont{Kono}}, in
  \emph{\bibinfo{booktitle}{Methods in Materials Research}}, edited by
  \bibinfo{editor}{\bibfnamefont{E.~N.} \bibnamefont{Kaufmann}} {\it et~al}.,
  %\bibinfo{editor}{\bibfnamefont{R.}~\bibnamefont{Abbaschian}},
  %\bibinfo{editor}{\bibfnamefont{A.}~\bibnamefont{Bocarsly}},
  %\bibinfo{editor}{\bibfnamefont{C.-L.} \bibnamefont{Chien}},
  %\bibinfo{editor}{\bibfnamefont{D.}~\bibnamefont{Dollimore}},
  %\bibinfo{editor}{\bibfnamefont{B.}~\bibnamefont{Doyle}},
  %\bibinfo{editor}{\bibfnamefont{A.}~\bibnamefont{Goldman}},
  %\bibinfo{editor}{\bibfnamefont{R.}~\bibnamefont{Gronsky}},
  %\bibinfo{editor}{\bibfnamefont{S.}~\bibnamefont{Pearton}}, \bibnamefont{and}
  %\bibinfo{editor}{\bibfnamefont{J.}~\bibnamefont{Sanchez}}
  (\bibinfo{publisher}{John Wiley \& Sons}, \bibinfo{address}{New York},
  \bibinfo{year}{2001}), chap. \bibinfo{chapter}{9b.2}.

\bibitem[{\citenamefont{Wang et~al.}(2007)\citenamefont{Wang, Hilton, Ren,
  Mittleman, Kono, and Reno}}]{WangetAl07OL}
\bibinfo{author}{\bibfnamefont{X.}~\bibnamefont{Wang}} {\it et~al}.,
  %\bibinfo{author}{\bibfnamefont{D.~J.} \bibnamefont{Hilton}},
  %\bibinfo{author}{\bibfnamefont{L.}~\bibnamefont{Ren}},
  %\bibinfo{author}{\bibfnamefont{D.~M.} \bibnamefont{Mittleman}},
  %\bibinfo{author}{\bibfnamefont{J.}~\bibnamefont{Kono}}, \bibnamefont{and}
  %\bibinfo{author}{\bibfnamefont{J.~L.} \bibnamefont{Reno}},
  \bibinfo{journal}{Opt. Lett.} \textbf{\bibinfo{volume}{32}},
  \bibinfo{pages}{1845} (\bibinfo{year}{2007});
%
%\bibitem[{\citenamefont{Wang et~al.}(2010)\citenamefont{Wang, Hilton, Reno,
  %Mittleman, and Kono}}]{WangetAl10OE}
%\bibinfo{author}{\bibfnamefont{X.}~\bibnamefont{Wang}},
  %\bibinfo{author}{\bibfnamefont{D.~J.} \bibnamefont{Hilton}},
  %\bibinfo{author}{\bibfnamefont{J.~L.} \bibnamefont{Reno}},
  %\bibinfo{author}{\bibfnamefont{D.~M.} \bibnamefont{Mittleman}},
  %\bibnamefont{and} \bibinfo{author}{\bibfnamefont{J.}~\bibnamefont{Kono}},
  \bibinfo{journal}{Opt. Exp.} \textbf{\bibinfo{volume}{18}},
  \bibinfo{pages}{12354} (\bibinfo{year}{2010}).
  
\bibitem{Suppl}
    See supplementary material for details of the experimental and theoretical methods used.

\bibitem[{\citenamefont{Thomson et~al.}(2007)\citenamefont{Thomson, Kre{\ss},
  L{\"o}ffler, and Roskos}}]{ThomsonetAl07LPR}
\bibinfo{author}{\bibfnamefont{M.~D.} \bibnamefont{Thomson}},
  \bibinfo{author}{\bibfnamefont{M.}~\bibnamefont{Kre{\ss}}},
  \bibinfo{author}{\bibfnamefont{T.}~\bibnamefont{L{\"o}ffler}},
  \bibnamefont{and} \bibinfo{author}{\bibfnamefont{H.~G.}
  \bibnamefont{Roskos}}, 
  \bibinfo{journal}{Laser \& Photon. Rev.}
  \textbf{\bibinfo{volume}{1}}, \bibinfo{pages}{349} (\bibinfo{year}{2007}).

\bibitem[{\citenamefont{Karpowicz et~al.}(2009)\citenamefont{Karpowicz, Lu, and
  Zhang}}]{KarpowiczetAl09JMO}
\bibinfo{author}{\bibfnamefont{N.}~\bibnamefont{Karpowicz}} {\it et~al}.,
  %\bibinfo{author}{\bibfnamefont{X.}~\bibnamefont{Lu}}, \bibnamefont{and}
  %\bibinfo{author}{\bibfnamefont{X.-C.} \bibnamefont{Zhang}},
  \bibinfo{journal}{J. Mod. Opt.} \textbf{\bibinfo{volume}{56}}
  (\bibinfo{year}{2009}).

\bibitem[{\citenamefont{Planken et~al.}(2001)\citenamefont{Planken, Nienhuys,
  Bakker, and Wenckebach}}]{PlankenetAl01JOSAB}
\bibinfo{author}{\bibfnamefont{P.~C.~M.} \bibnamefont{Planken}},
  \bibinfo{author}{\bibfnamefont{H.-K.} \bibnamefont{Nienhuys}},
  \bibinfo{author}{\bibfnamefont{H.~J.} \bibnamefont{Bakker}},
  \bibnamefont{and}
  \bibinfo{author}{\bibfnamefont{T.}~\bibnamefont{Wenckebach}},
  \bibinfo{journal}{J. Opt. Soc. Am. B} \textbf{\bibinfo{volume}{18}},
  \bibinfo{pages}{313} (\bibinfo{year}{2001}).

\bibitem[{\citenamefont{Palik and Furdyna}(1970)}]{PalikFurdyna70RPP}
\bibinfo{author}{\bibfnamefont{E.~D.} \bibnamefont{Palik}} \bibnamefont{and}
  \bibinfo{author}{\bibfnamefont{J.~K.} \bibnamefont{Furdyna}},
  \bibinfo{journal}{Rep. Prog. Phys.} \textbf{\bibinfo{volume}{33}},
  \bibinfo{pages}{1193} (\bibinfo{year}{1970}).

\bibitem[{\citenamefont{Boyd}(2008)}]{Boyd08Book}
\bibinfo{author}{\bibfnamefont{R.~W.} \bibnamefont{Boyd}},
  \emph{\bibinfo{title}{Nonlinear Optics}} (\bibinfo{publisher}{Academic
  Press}, \bibinfo{address}{New York}, \bibinfo{year}{2008}).

\bibitem[{\citenamefont{Karadimitriou et~al.}(2010)\citenamefont{Karadimitriou,
  Kavousanaki, Perakis, and Dani}}]{KaradimitriouetAl10PRB}
\bibinfo{author}{\bibfnamefont{M.~E.} \bibnamefont{Karadimitriou}},
  \bibinfo{author}{\bibfnamefont{E.~G.} \bibnamefont{Kavousanaki}},
  \bibinfo{author}{\bibfnamefont{I.~E.} \bibnamefont{Perakis}},
  \bibnamefont{and} \bibinfo{author}{\bibfnamefont{K.~M.} \bibnamefont{Dani}},
  \bibinfo{journal}{Phys. Rev. B} \textbf{\bibinfo{volume}{82}},
  \bibinfo{pages}{165313} (\bibinfo{year}{2010}).

\bibitem[{\citenamefont{Broocks et~al.}(2008)\citenamefont{Broocks, Su,
  Schr{\"o}ter, Heyn, Heitmann, Wegscheider, Apalkov, Chakraborty, Perakis, and
  Sch{\"u}ller}}]{BroocksetAl08PSSB}
\bibinfo{author}{\bibfnamefont{K.~B.} \bibnamefont{Broocks}} {\it et~al}.,
  %\bibinfo{author}{\bibfnamefont{B.}~\bibnamefont{Su}},
  %\bibinfo{author}{\bibfnamefont{P.}~\bibnamefont{Schr{\"o}ter}},
  %\bibinfo{author}{\bibfnamefont{C.~H.} \bibnamefont{Heyn}},
  %\bibinfo{author}{\bibfnamefont{D.}~\bibnamefont{Heitmann}},
  %\bibinfo{author}{\bibfnamefont{W.}~\bibnamefont{Wegscheider}},
  %\bibinfo{author}{\bibfnamefont{V.~M.} \bibnamefont{Apalkov}},
  %\bibinfo{author}{\bibfnamefont{T.}~\bibnamefont{Chakraborty}},
  %\bibinfo{author}{\bibfnamefont{I.~E.} \bibnamefont{Perakis}},
  %\bibnamefont{and}
  %\bibinfo{author}{\bibfnamefont{C.}~\bibnamefont{Sch{\"u}ller}},
  \bibinfo{journal}{Phys. Stat. Solid. B} \textbf{\bibinfo{volume}{245}},
  \bibinfo{pages}{321} (\bibinfo{year}{2008}).

\end{thebibliography}

\end{document}